\documentclass[final,5p,times,twocolumn]{elsarticle}

\usepackage{graphicx}
\usepackage{amssymb}
\usepackage{float}
\usepackage[switch]{lineno}

\biboptions{numbers,sort&compress}

\usepackage{svg}
\usepackage{amsmath}
\usepackage{dblfloatfix} 
\usepackage{wrapfig} 
\usepackage{nomencl}
\makenomenclature
\usepackage{afterpage}

\usepackage{etoolbox}
\renewcommand\nomgroup[1]{%
  \item[\bfseries
  \ifstrequal{#1}{A}{Subscript}{%
  \ifstrequal{#1}{B}{Abbreviations}{%
  \ifstrequal{#1}{C}{Variable}{}}}%
]}

\usepackage{adjustbox}  
\usepackage{multirow}   
\usepackage{url} 
\usepackage{eurosym}
\usepackage{amsfonts} 
\usepackage{dblfloatfix} 
\usepackage{hyperref} 

\usepackage[colorinlistoftodos]{todonotes}
\usepackage[utf8]{inputenc} 
\usepackage{multirow}
\usepackage{array}
\newcommand\Tstrut{\rule{0pt}{3ex}}         
\newcommand\Bstrut{\rule[-1.5ex]{0pt}{0pt}}   

\usepackage[nolist]{acronym} 			

\journal{TBD}

\begin{document}

\begin{frontmatter}


\title{Reducing energy system model distortions from unintended storage cycling through variable costs}










\newcommand{\orcidauthorA}{0000-0002-4390-0063}
\newcommand{\orcidauthorB}{0000-0002-3494-0469} 



\author[First]{Maximilian Parzen \corref{cor1}\fnref{label2}}
\ead{max.parzen@ed.ac.uk}

\author[Second,Third]{Martin Kittel}
\author[First]{Daniel Friedrich}
\author[First]{Aristides Kiprakis}

\cortext[cor1]{Corresponding author.}

\address[First]{University of Edinburgh, Institute for Energy Systems, EH9 3DW Edinburgh, United Kingdom}
\address[Second]{German Institute for Economic Research, Mohrenstra{\ss}e 58, 10117 Berlin, Germany}
\address[Third]{Technical University Berlin, Department of Energy Systems, Einsteinufer 25 (TA 8), 10587 Berlin, Germany}

\begin{abstract}

Energy model distortions may unknowingly appear and represent unrealistic and non-physical effects that can mislead optimal model decisions. A prominent misleading effect is unintended storage cycling that was observed in previous literature by simultaneous charging and discharging of the same storage in 12 of 18 energy models. Especially for popular net-zero energy model scenarios, unintended storage cycling can cause high distortions and, hence, urges for its removal. Methods to remove such misleading effects exist, but are not computational efficient and sometimes not effective such as MILP formulations. Other techniques are successful, but only if renewable energy target constraints exist. This paper explores how to eradicate unintended storage cycling for models without renewable target constraints by correctly setting variable costs of relevant system components.
We find through 124 simulations that determining appropriate levels of variable costs depends on the solver accuracy used for the optimization. If set too loose, the solver prevents the removal of unintended storage cycling. We further find that reliable data for variable costs in energy modelling needs to be improved and provide a list of recommended model inputs as well as a minimum variable cost threshold that can significantly reduce the magnitude and likeliness of unintended storage cycling.
Finally, our results suggest that variable cost additives may remove other known unintended energy cycling effects, such as unintended line cycling or sector cycling.

\end{abstract}


\begin{highlights}
\item Unspecified costs of energy dissipation options may cause excessive storage use 
\item Observed operational distortions up to 23\% for energy storage and 5\% for renewables 
\item Novelty is that variable cost additives can avoid such distortions 
\item Effective variable cost additives depends on solver accuracy and component allocation 
\item A list of variable cost suggestions is provided that may alleviate distortions 
\end{highlights}


\end{frontmatter}

\begin{acronym}
  \acro{HVAC}{high voltage alternative current}
  \acro{HVDC}{high voltage direct current}
  \acro{USC}{unintended storage cycling}
  \acro{VRE}{variable renewable energy sources}
  \acro{FLH}{full load hours}
  \acro{CSP}{concentrated solar power}
  \acro{PV}{photovoltaic}
  \acro{IPM}{interior point method}
  \acro{H2}{hydrogen}
  \acro{GHG}{greenhouse gas}
\end{acronym}



\section{Introduction}\label{S:Introduction}

Energy system models are mathematical models used to investigate possible pathways for decarbonising our energy systems, in many cases minimising total system costs \cite{Ringkjb2018ARenewables}. They provide insights on optimal dispatch and investment patterns in the short- and long-term, which guide energy technology design decisions \cite{Parzen2022BeyondSystems} and support the decision-making of governments, grid operators, energy system planners, manufacturers, and researchers. However, if not carefully used, such models can mislead decision-making. 

\begin{figure}[h!]
\centering
\includegraphics[trim={6cm 6.92cm 10cm 7.1cm},clip,width=0.48\textwidth]{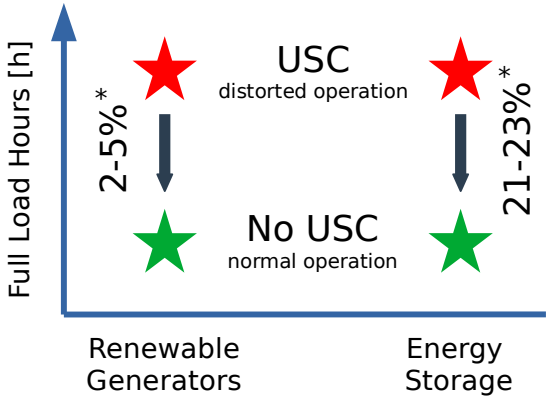}
\caption{Exemplary impact of scenarios with and without \ac{USC} on storage and renewable generation. The \ac{USC} effect increases the operation of close to zero variable cost assets. Results from a numerical analysis conducted in this study, marked with an  asterisk ($^*$), show \ac{FLH} differences of up to 23\% for energy storage and 5\% for renewable assets in a 100\% renewable energy system scenario.}
\label{fig:USC effect intro}
\end{figure}

\begin{figure}[h!]
\centering
\includegraphics[trim={0.5cm 1cm 0.5cm 1cm},clip,width=0.48\textwidth]{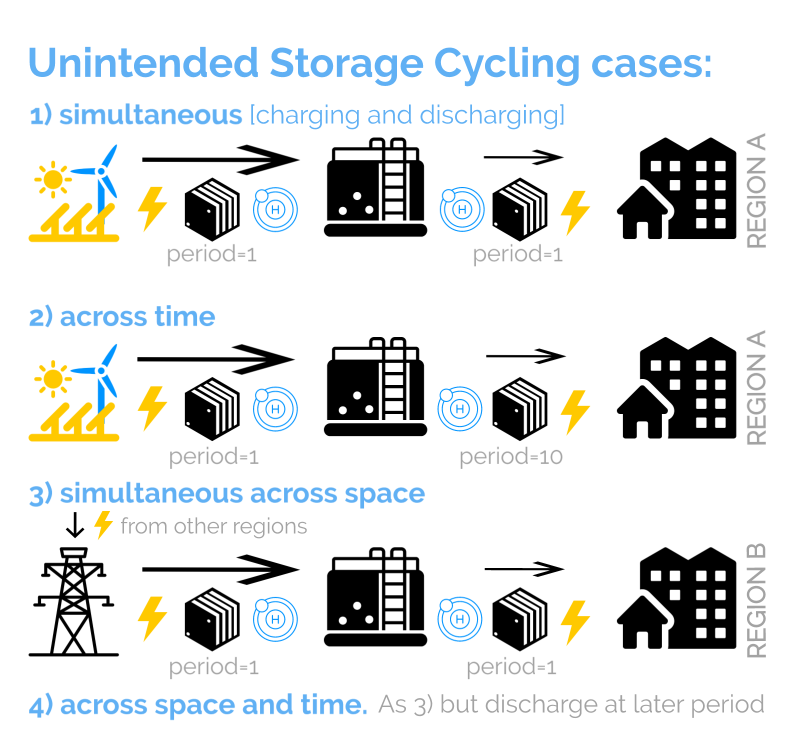}
\caption{Unintended storage cycling cases in a hydrogen storage example. The left stack represents the electrolyser, the right one, the fuel cell which converts electricity to hydrogen and vice versa, respectively. The arrow size above the storage components reduces to indicate an efficiency drop. Under renewable energy surplus (variable cost = 0), excess energy is removed from the system by storage cycling instead of renewable curtailment. Thereby, \ac{USC} may occur over space and time if no constraint prohibits this artifact.}
\label{fig:USC classification intro}
\end{figure}

One model artifact distorting optimal model results is \ac{USC} \cite{Kittel2021RenewableModeling}, which has been observed in 12 of 18 well-established energy system models as reviewed by Kittel and Schill (2022) \cite{Kittel2021RenewableModeling}. The effect impacts storage use. Instead of curtailing \ac{VRE} surplus, the excess electricity is converted into unintended storage losses by simultaneous charging and discharging of the same storage capacity. The consequence of this behaviour are distortions of optimal model outcomes. For example, energy storage or renewable generators may have significantly more \ac{FLH} in a scenario with, compared to one without \ac{USC} (Figure \ref{fig:USC effect intro}), signalling deceptively more intensive operation. Most importantly, \ac{USC} is technically infeasible for some storage technologies such as the widely deployed lithium-ion battery storage that can either charge or discharge but not both simultaneously. Hence, the effect urges its removal. \ac{USC} may also manifest across space and time (Figure \ref{fig:USC classification intro}). For instance, \ac{USC} across space may occur in multi-regional model settings through simultaneous charging in one region and discharging in another region for the only purpose to dissipate renewable surplus energy instead of curtailing it, notably in the absence of transmission costs. Similar, \ac{USC} across time represents unintended simultaneous charging and discharging cycles across multiple periods \cite{Kittel2021RenewableModeling}. \ac{USC} is not bound to the same spatial and temporal cycling moments, which aligns with the non-guaranteed operational uniqueness in scenarios with multiple storage assets \cite{Grubel2020OnADMM}.

\ac{USC} is not the only such misleading effect. There is a group of related effects classified under the term unintended energy losses, which arise in a cyclic manner \cite{Kittel2021RenewableModeling}. Unintended, because they distort optimization results. Cyclic, as they occur in the energy systems wherever efficiency losses are present between energy components that cycle energy by charging and discharging, sending and receiving, or converting and re-converting operations.

The literature on unintended energy losses is limited. State-of-the-art guidance on best-practice energy system modelling probably unintentionally ignores these artefacts \cite{DeCarolis2017FormalizingModelling, Bistline2020EnergyNeeds}, while others do this intentionally \cite{Neumann2019HeuristicsModels}. Attempts to remove unintended energy losses exist. For instance, Yang et al. in \cite{Yang2018AOPF} penalise active power losses, which can be too simplified and distort the model outputs by falsely allocating the costs. Other studies create a high number of linear constraints and reformulate the problem to a quadratic or mixed-integer problem \cite{Kittel2021RenewableModeling, Fortenbacher2019Linear/QuadraticApproximations}, which is harder to solve and requires more computational requirements. Moreover, these mixed-integer approaches cannot guarantee the full \ac{USC} removal, since they only prohibit simultaneous charging and discharging with binary variables, which fails with \ac{USC} across space and time. 

Kittel and Schill (2022) \cite{Kittel2021RenewableModeling} investigate \ac{USC} arising in models that are constrained by a binding renewable energy target. In these models, \ac{USC} causes an increase of \ac{VRE} generation, which can be realised without additional renewable capacity installations. Thus, the renewable energy target can be achieved with less \ac{VRE} capacity at lower costs. Yet, it does not serve demand, which requires additional generation from other dispatchable technologies. Here, \ac{USC} flaws both optimal dispatch and investment decisions. To remove \ac{USC}, the renewable energy constraint has to include these unintended energy losses, preventing the cost-minimizing conversion of intended \ac{VRE} curtailment into unintended energy losses. 

However, the solution presented in \cite{Kittel2021RenewableModeling} is not applicable to model formulations \textit{without} binding renewable targets \cite{Henke2022TheModelling, Pavicevic2021Water-energyModel, Victoria2020EarlyOff, Brown2018PyPSA:Analysis}. In such models, \ac{USC} is caused by a different mechanism: It may arise if different options to dissipate unused energy from the system are available and come either at zero or at the same costs. Then, the optimization becomes indifferent with regard to energy dissipation options. \ac{VRE} surplus energy can either be curtailed, or erased from the system via unintended energy losses from \ac{USC}, distorting optimal dispatch results (see Section \ref{sec: detect usc} and \ref{Sec:removing usc}). 

This paper contributes to the existing literature in several aspects:  Firstly, we provide a method for the removal of \ac{USC} in linear models \textit{without} binding renewable energy targets. We show that the removal can be achieved by a deliberate setting of variable costs of affected system components. Unlike MILP-based approaches, this solution keeps the problem formulation linear, making it more effective and efficient. Secondly, we mathematically formalise how \ac{USC} occurs. Thirdly, we explore the impact of \ac{USC} and its removal on operational and investment decisions as well as total system cost in a decarbonised German energy system model. To this end, we demonstrate how variable costs, their magnitude and allocation can remove \ac{USC}, while also giving new insights on the role of the solver accuracy. Finally, we provide a reviewed list of variable cost inputs, which may guide the removal of \ac{USC} in other energy models.

\section{Methods}\label{S:Methodology}

\subsection{Model formulation}\label{Sec: model formulation}

The occurrence of \ac{USC} depends on the model formulation \cite{Kittel2021RenewableModeling}. This section introduces a general formulation of investment and dispatch optimization problems for energy system models that abstain from binding renewable energy target constraints. The objective of such models is to minimise the total system costs, comprised of annualised capital and operational expenditures. Capital expenditures include capacity-related, long-term investment costs $c$ at location $i$ for generator $G_{i,r}$ of technology $r$, storage energy capacity $H_{i,s}^{store}$, charging capacity $H^+_{i,s}$ and discharging capacity $H^-_{i,s}$ of technology $s$ and transmission line $F_{l}$. Operational expenditures include energy-related variable cost $o$ for generation $g_{i,r,t}$ and storage charging $h^+_{i,r,t}$ and discharging $h^-_{i,r,t}$, as well as energy-level related storage cost $e_{i,s,t}$. Thereby, the operation depends on the time steps $t$ that are weighted by duration $w_{t}$ that sums up to one year  $\sum_{t=1}^{T} w_{t}= 365days*24h = 8760h$.

\begin{equation}
\begin{aligned}
\min_{G,H,F,g,h,e} \quad & \left(\text{Total System Cost}\right) = \\
\min_{G,H,F,g,h,e} \quad &   
\Bigg[ \sum_{i,r}(c_{i,r} \cdot G_{i,r}) 
+
\sum_{l}(c_{l} \cdot F_{l}) \\
\quad & 
+ \sum_{i,s}(c^{store}_{i,s} \cdot H_{i,s}^{store} + c^{-}_{i,s} \cdot H^-_{i,s} + c^+_{i,s} \cdot H^+_{i,s}) \\
\quad & 
+ \sum_{i,r,t}(o_{i,r} \cdot g_{i,r,t} \cdot w_t) + 
\sum_{i,s,t}\big((o^+_{i,s} \cdot h^+_{i,s,t} + o^-_{i,s} \cdot h^-_{i,s,t}) \cdot w_t\big) \\
\quad & 
+ \sum_{i,s,t}(o^{store}_{i,s} \cdot e_{i,s,t} \cdot w_t) \Bigg] \\
\end{aligned}
\end{equation}

The objective function can be subject to multiple linear constraints, an example is laid out in more detail in \cite{Neumann2021TheModel, Horsch2018PyPSA-Eur:System}, leading to a convex linear program. A convex linearly formulated problem has a unique objective value with sometimes multiple non-unique operational solutions. For instance, when various ways exist to distribute the storage and generation operation temporally and spatially, in the absence of spatial limitations (i.e., available transmission capacity) or temporal limitations (i.e., available energy store capacity) for consecutive time steps. This is also the reason that \ac{USC} can occur over space and time, as mentioned in the introduction, and arise in a non-unique pattern as long as it does not break any constraint or lead to higher costs.

Further, constraints in the energy model include:

\begin{itemize}
    \item nodal power balance constraints guaranteeing that supply equals demand at all times,
    \item linearised power flow constraints modelling the physicality of power transmission (Kirchhoff's Voltage and Current Law),
    \item hourly solar and wind resource availability constraints limiting the renewable generation potential based on reanalysis weather data,
    \item land-use constraints, restricting the renewable capacity expansion based on environmental protection areas, land use coverage, and distance criteria, and finally,
    \item emission constraints introducing a limit of $CO_2$ equivalent \ac{GHG} emissions.
\end{itemize}

Storage charging $h_{i,s,t}^+$ and discharging $h_{i,s,t}^-$ are both positive variables and limited by the installed capacity $H_{i,s,t}^+$ and $H_{i,s,t}^-$.

\begin{equation}
\begin{aligned}
0 \leq h_{i,s,t}^+ \leq H_{i,s}^+ \quad \forall i,s,t 
\end{aligned}
\end{equation}

\begin{equation}
\begin{aligned}
0 \leq h_{i,s,t}^- \leq H_{i,s}^- \quad \forall i,s,t
\end{aligned}
\end{equation}

This formulation keeps the feasible solution space convex.

The storage energy level $e_{i,s,t}$ is the result of a balance between energy inflow, outflow and self-consumption. Additional to directed charging and discharging with its respective efficiencies $\eta_{i,s,+}$ and $\eta_{i,s,-}$, natural inflow $h_{i,s,t}^{inflow}$, spillage $h_{i,s,t}^{spillage}$ as well as standing storage losses that reduces the storage energy content of the previous time step by a factor of $\eta_{i,s,+}$ are considered.

\begin{equation}
\begin{aligned}
e_{i,s,t} \quad & = \eta_{i,s,+} \cdot e_{i,s,t-1} + \eta_{i,s,+} \cdot w_t \cdot h_{i,s,t}^+ - \eta_{i,s,-}^{-1} \cdot w_t \cdot h_{i,s,t}^- \\
\quad & + w_t \cdot h_{i,s,t}^{inflow} - w_t \cdot h_{i,s,t}^{spillage} \quad \forall i,s,t
\end{aligned}
\end{equation}

The amount of energy that can be stored is limited by the energy capacity of the installed store unit $H_{i,s}^{store}$ [MWh], which allows independent storage component scaling. 

\begin{equation}
\begin{aligned}
0 \leq e_{i,s,t} \leq H_{i,s}^{store} \quad \forall i,s,t 
\end{aligned}
\end{equation}

To fix the storage technology design, a technology-specific energy to discharging power ratio $\overline{T}_s$ can be multiplied with the capacity of the discharging unit $H_{i,s}^-$ 

\begin{equation}
\begin{aligned}
0 \leq e_{i,s,t} \leq \overline{T}_s \cdot H_{i,s}^- \quad \forall i,s,t
\end{aligned}
\end{equation}

\noindent to define the upper energy limit per installed storage.

Finally, energy storage units are assumed to be cyclic, i.e., the state of charge at the first and last period of the optimization period $T$ (i.e. 1 year) must be equal:

\begin{equation}
\begin{aligned}
e_{i,s,0} = e_{i,s,T} \quad \forall i,s
\end{aligned}
\end{equation}

This cyclic definition is not mandatory but helps with the comparability of model results. It further avoids the free use of storage energy endowment, meaning that the model could prefer to start with a higher and end with a lower storage level to save costs. 

\subsection{Detecting unintended storage cycling occurrence}\label{sec: detect usc}

\ac{USC} can be identified by analysing storage charging and discharging patterns. A straightforward approach is to count the occurrence of simultaneous charging and discharging over the optimization horizon which we use in later parts of the study. Here, \ac{USC} may occur under three cases in energy systems with energy storage\cite{Kittel2021RenewableModeling}: During effective charging, effective discharging, or in an idle energy state with effective net-zero charging.

The storage charging power $h_{i,s,t}^{+}$ describes the power provision from the grid to the charging component. If reduced by the charging efficiency $\eta_{i,s,+}$, it results in storage charging power $h_{i,s,t,store}^{+}$ that increases the storage energy level over time.

\begin{equation}
\begin{aligned}
h_{i,s,t,store}^+ = \eta_{i,s,+} \cdot h_{i,s,t}^+
\end{aligned}
\end{equation}

Likewise, store discharging power $h_{i,s,t, store}^-$ describes the power provision from the storage that reduces the storage energy level over time. If reduced by the discharging efficiency $\eta_{i,s,-}$, it results in the storage discharging power $h_{i,s,t}^-$ that provides power to the grid.

\begin{equation}
\begin{aligned}
h_{i,s,t,store}^- = \frac{h_{i,s,t}^-}{\eta_{i,s,-}}
\end{aligned}
\end{equation}

The first case \ac{USC} occurs under effective charging ${USC}_{i,s,t}^{+}$, which increases the storage energy level over time:

\begin{equation}
\begin{aligned}
if \quad & h_{i,s,t,store}^+ > h_{i,s,t,store}^- \quad  \text{and} \quad h_{i,s,t,store}^- > 0 \colon  \\
& \text{USC}_{i,s,t}^{+} = \text{true} 
\end{aligned}
\end{equation}

The second case occurs under effective discharging ${USC}_{i,s,t}^{-}$, which decreases the storage energy level over time:

\begin{equation}
\begin{aligned}
if \quad & h_{i,s,t,store}^+ < h_{i,s,t,store}^- \quad  \text{and} \quad h_{i,s,t,store}^+ > 0 \colon  \\
& \text{USC}_{i,s,t}^{-} = \text{true} 
\end{aligned}
\end{equation}

The third case appears under non-zero equal charging and discharging ${USC}_{i,s,t}^{=}$, or idle energy state, which keeps the storage energy level over time constant (neglecting standing losses):

\begin{equation}
\begin{aligned}
if & \quad h_{i,s,t,store}^+ = h_{i,s,t,store}^- \quad  \text{and} \quad h_{i,s,t,store}^+ > 0 \colon \\
& \quad \text{USC}_{i,s,t}^{=} = \text{true} 
\end{aligned}
\end{equation}


\begin{table*}[t!]
\caption{Model input assumptions.}
\label{Table. Model Input}
  \centering
  \resizebox{\textwidth}{!}{
  \begin{tabular}{lll lll ll}
    \hline
    Technology $^a$ \Tstrut\Bstrut & Investment & Fixed O\&M & variable cost $^b$ & Lifetime & Efficiency & Source\\
    & [\EUR{}/kW] & [\EUR{}/kW/a] & [\EUR{}/MWh] & [a] & [-] &  \Bstrut \\
    \hline
    onshore wind \Tstrut & 1040 & 25 & variable & 30 & 1 & DEA \cite{DanishEnergyAgencyDEA2019TechnologyData}\\
    offshore wind (\acs{HVAC} connected) & 1890 & 44 & variable & 30 & 1 & DEA \cite{DanishEnergyAgencyDEA2019TechnologyData} \\
    offshore (\acs{HVDC} connected) & 2040 & 47 & variable & 30 & 1 & DEA \cite{DanishEnergyAgencyDEA2019TechnologyData} \\
    \ac{PV} & 600 & 25 & variable & 25 & 1 & Schröder et al. \cite{Schroder2013Prospective2050} \\
    hydrogen electrolyser & 350 & 14 & variable & 25 & 0.8 & Budischak et al. \cite{Budischak2013Cost-minimizedTime} \\
    hydrogen storage tank & 8.44 \EUR/MWh & - & variable & 20 & 1 & Budischak et al. \cite{Budischak2013Cost-minimizedTime} \\
    hydrogen fuel cell & 339 & 10 & variable & 20 & 0.58 & Budischak et al. \cite{Budischak2013Cost-minimizedTime} \\
    transmission (submarine) & 2000 \EUR/MWkm & 2\%/a & 0 & 40 & 1 & Hagspiel et al. \cite{Hagspiel2014Cost-optimalCoupling} \\
    transmission (overhead) & 400 \EUR/MWkm & 2\%/a & 0 & 40 & 1 & Hagspiel et al. \cite{Hagspiel2014Cost-optimalCoupling} \Bstrut \\
    \hline

\multicolumn{4}{l}{$^a$ All technologies include a discount rate of 7\%. \Tstrut } \\
\multicolumn{4}{l}{$^b$ 'Variable' means set according to scenarios.} \\
\multicolumn{6}{l}{$^c$ Unconstrained energy storage sizing and not fixed to a specific energy to power ratio.}
  \end{tabular}
  } 
  \label{tab:1}
\end{table*}




\subsection{Removing unintended storage cycling by variable cost additives}\label{Sec:removing usc}



We define variable cost additives as not necessarily true observed variable costs, but more generally as assumed additional costs components.

Suppose an optimization found a least-cost total system architecture. Then the system operation may adapt any value as long as it does not lead to more cost (left side of Equation (\ref{usceq}) and (\ref{nousceq})) and does not break constraints such as demand is equal supply (described in \ref{Sec: model formulation}).

Further suppose, the energy system contains a renewable energy surplus at a time step, and variable operational cost $o_{i,s/r}$ of storage and renewables are assumed as zero, then: 

\begin{equation}\label{usceq}
\begin{aligned}
0 & \quad = 
\underbrace{o_{i,r}}_\text{0} 
\cdot g_{i,r,t}^* + 
\underbrace{o_{i,s}^{+}}_\text{0} \cdot h_{i,s,t}^{*,+}
+
\underbrace{o_{i,s}^{-}}_\text{0} \cdot h_{i,s,t}^{*,-}
+
\underbrace{o_{i,s}^{store}}_\text{0} \cdot
\triangle e_{i,s,t}^* \\
\end{aligned}
\end{equation}  

These zero costs may lead to a situation where surplus generation is fed into the grid rather than curtailed. However, to guarantee the energy balance the extra surplus generation needs to be dissipated by \ac{USC} (indicated by $^*$), which leads to higher storage usage.

In contrast, in case costs exist for either generation or storage operation,

\begin{equation}\label{nousceq}
\begin{aligned}
0 & \quad = 
o_{i,r}
\cdot 
\underbrace{g_{i,r,t}}_\text{0} + 
o_{i,s}^{+} \cdot 
\underbrace{h_{i,s,t}^{+}}_\text{0}
+
o_{i,s}^{-} \cdot \underbrace{h_{i,s,t}^{-}}_\text{0}
+
o_{i,s}^{store}
\cdot
\underbrace{\triangle e_{i,s,t}}_\text{0} \\
\end{aligned}
\end{equation} 

every additional operation of variable renewable generators or storage is prevented in the first place, avoiding \ac{USC}.

Equations (\ref{usceq}) and (\ref{nousceq}) illustrate that a system with \ac{USC} (indicated by $^*$) has components with higher operating hours than one without,

\begin{equation}
\begin{aligned}
& \quad \sum_{t=1}^{T} g_{i,r,t}^* \geq \sum_{t=1}^{T} g_{i,r,t} \\
\end{aligned}
\end{equation}

\begin{equation}
\begin{aligned}
& \quad \sum_{t=1}^{T} h_{i,s,t}^{*,+/-} \geq \sum_{t=1}^{T} h_{i,s,t}^{+/-} \\
\end{aligned}
\end{equation}

\begin{equation}
\begin{aligned}
& \quad \sum_{t=1}^{T} e_{i,s,t}^{*} \geq \sum_{t=1}^{T} e_{i,s,t} \\
\end{aligned}
\end{equation}

caused by energy dissipation through excessive storage use rather than curtailing renewable surplus. 

In summary, to remove \ac{USC}, a situation with \ac{USC} must become more expensive than one without because, fundamentally, the objective function aims to minimise cost. One approach is to add variable cost $o_{i,r} > 0$ to the generation dispatch. Another one is to add variable costs $o_{i,s} > 0$ to any or all energy storage components such as charger, store or discharger. Both such variable cost additives penalise any extra operation of generators or storage units caused by \ac{USC} energy dissipation even across space and time. Nevertheless, since variable costs not only penalise \ac{USC} but the operation of these units, they need to be carefully chosen.

\subsection{Numerical implementation and data}

We use PyPSA-Eur \cite{Horsch2018PyPSA-Eur:System} as a numerical implementation for the model defined in Section \ref{Sec: model formulation} to explore the occurrence and amplitude of \ac{USC}. PyPSA-Eur is a European power system model, representative of energy models that abstain from binding renewable energy targets. We apply the model to a stylised setting parameterised to the German power sector for a 100\% \ac{GHG} emission reduction scenario. We limit the available set of technologies to solar \ac{PV}, onshore wind, offshore wind, as well as an \ac{H2} storage system consisting of an electrolyzer, a tank, and a fuel cell. We set the spatial resolution to 16 nodes within Germany. Offshore wind power plants may be connected via \acf{HVAC}, or, in the case of sites far offshore, more costly \acf{HVDC} transmission lines. The model has perfect foresight and optimises with an hourly temporal resolution. Weather and load data stem from 2013. Hourly load data originates from the ENTSO-E Transparency platform and are distributed across the regions depending on NUTS3 based GDP data (see more in \cite{Horsch2018PyPSA-Eur:System}). All renewables and energy storage technologies are greenfield optimized, i.e., without considering the existing capital stock. State of charge of energy storage capacities is constrained to start and end with 100\%. The self-consumption of the \ac{H2} storage tank is assumed to be zero. The transmission network is based on the current network topology, considering also planned lines until 2030 from the ENTSO-E Ten Year Network Development Plan (TYNDP) 2018 \cite{ENTSO-E2018TYNDP:Plan}. Grid expansion is endogenous but limited to additional 25\% newly built lines for modelled target year to represent political hurdles of transmission expansion \cite{Neumann2021TheModel}. Table 1 lists relevant techno-economic assumptions. This stylised setting allows for demonstrating \ac{USC} in the context of energy models without binding renewable energy targets, and how it can be removed by a deliberate setting of variable costs.

\subsection{Experimental setup}

\begin{figure}[h]
\centering
\includegraphics[trim={0cm 0.0cm 0cm 0cm},clip,width=0.48\textwidth]{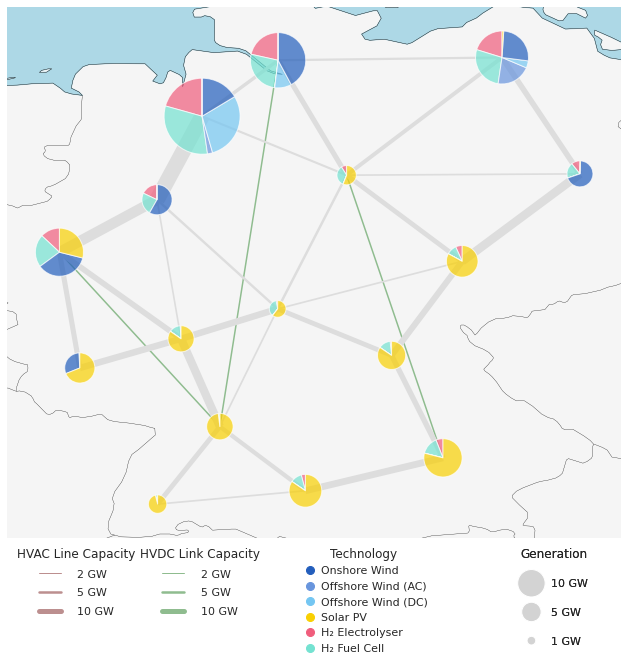}
\caption{Optimized generation and storage capacities in Germany for a 100\% \ac{GHG} emission reduction scenario.}
\label{fig:germany_map}
\end{figure}

To investigate the suggested method for removing \ac{USC}, in the base case scenario we set the variable cost (EUR/MWh) of the renewable generators, \ac{H2} electrolysers, \ac{H2} tanks, and \ac{H2} fuel cells to zero. We then define further scenarios varying \textit{ceteris paribus} the variable cost of one of these system components in the range $e \in \{0, 10^{-5}, 10^{-4}, 10^{-3}, 10^{-2}, 10^{-1}, 10^{0}, 10^{1}, 10^{2}, 10^{3}\}$. That is, in each scenario, the variable cost of one system component changes according to range $e$, while the others are kept constant at zero. Note that in the respective scenarios variable costs of all renewable generators are varied at once. Cost additives of 100 or 1000 \euro/MWh (10 or 100 ct/kWh) represent a demonstrative, non-realistic value for all included technologies that could be interpreted as falsely set variable costs.

The double-precision arithmetic limits the amount of numbers a computer can recognise. While optimization solvers are also influenced by the double-precision arithmetic, they additionally include tolerances to faster solve problems, which comes at the cost of accuracy. We vary the precision of two Gurobi solver parameters simultaneously yielding three scenarios: low, medium, and high accuracy. The first modified Gurobi parameter that impacts the precision is the $FeasibilityTol$ or primal feasibility tolerance, which requires all constraints to satisfy a specific tolerance to be feasible \cite{GurobiOptimization2020GurobiManual}. For instance, constraints such as $(a*x) <= b$ require to hold $(a*x)-b <= FeasibilityTol$ expanding the solution space. We vary this value by $e \in \{10^{-5},10^{-6},10^{-7} \}$ from low to high accuracy. Simultaneously, we vary the Gurobi parameter $BarConvTol$, which describes the barrier solver (also known as \ac{IPM}) termination tolerance as relative difference between the primal and dual objective values \cite{GurobiOptimization2020GurobiManual}. Given one solution space for an optimization problem this relative difference is also known as duality gap \cite{Boyd2004ConvexOptimization}, which IPM's reduces iteratively towards zero before the termination tolerance is reached. We vary this value by $e \in \{10^{-4},10^{-5},10^{-6} \}$, again, from low to high accuracy.

The computations to generate all results ($10*4*3=120$ scenarios) required 25.5 h for 1 CPU core with 8GB memory.

\section{Results and discussion}\label{S:Result and Discussion}



In the following subsections, we investigate the impact of \ac{USC} and its removal on model outcomes such as optimal dispatch, installed capacity, and total system cost. 

\subsection{Effects on operational optimization}

\begin{figure}[hbt!]
\centering
\includegraphics[trim={0cm 2.5cm 1cm 0cm},clip,width=0.48\textwidth]{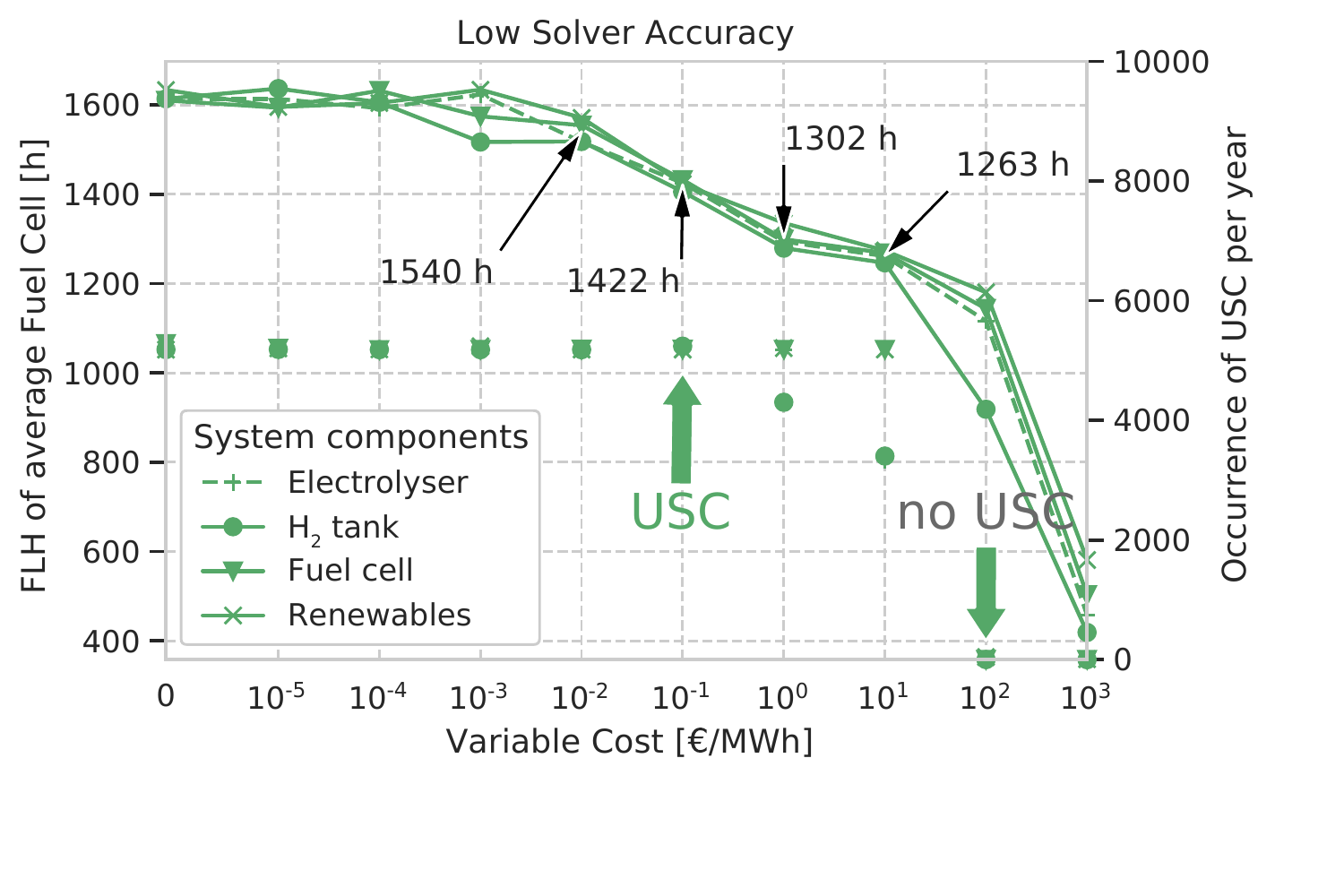}
\includegraphics[trim={0cm 2.5cm 1cm 0cm},clip,width=0.48\textwidth]{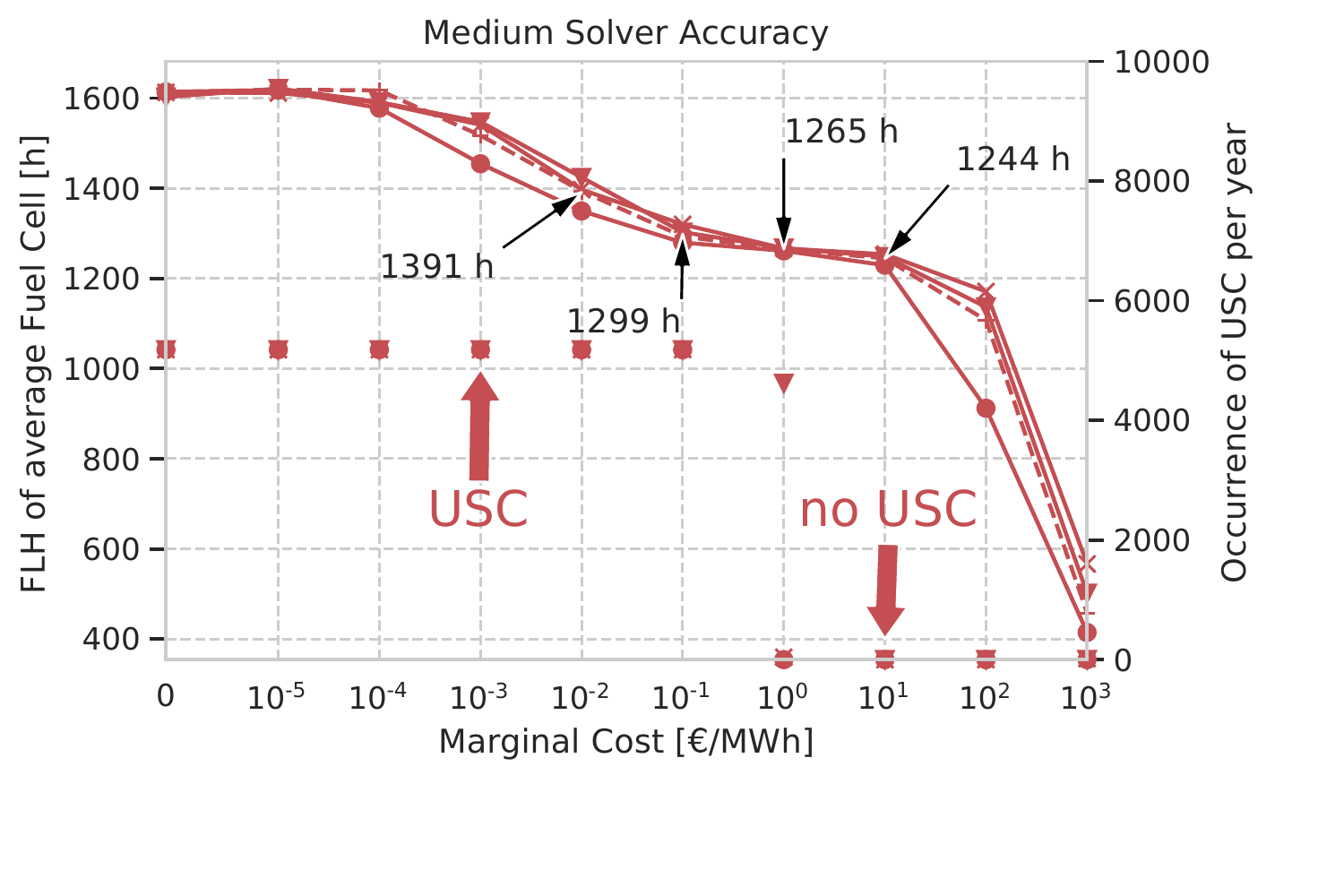}
\includegraphics[trim={0cm 3.5cm 1cm 0cm},clip,width=0.48\textwidth]{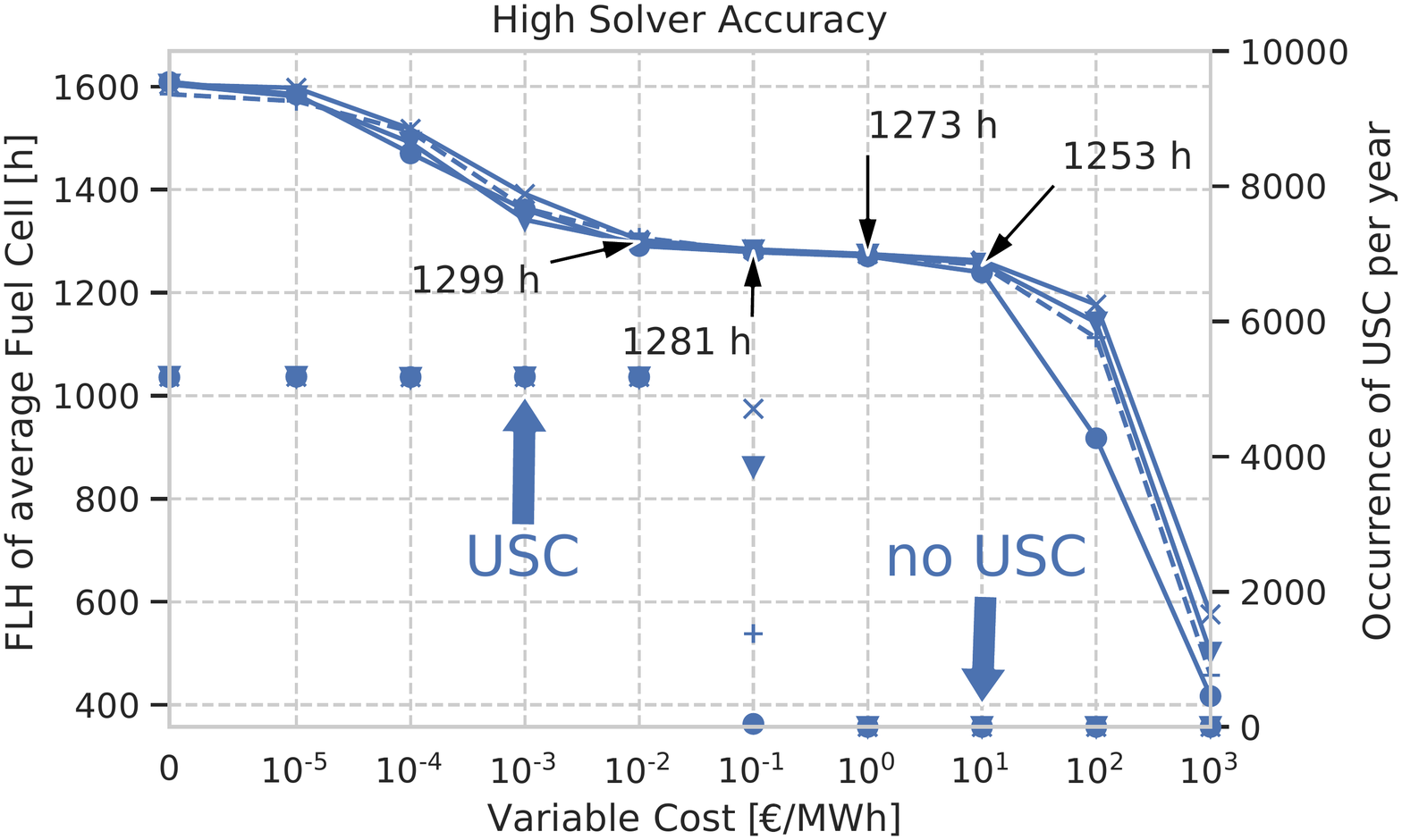}\caption{\ac{FLH} of the \ac{H2} fuel cell (lines, left ordinate) and occurrences of \ac{USC} (scatter plots, right ordinate) for three levels of solver accuracy (low, medium, high from top to bottom) across different levels of variable cost additives for renewable generators or \ac{H2} storage components (abscissa). \ac{FLH} are averaged across all nodes in Germany. Per scenario, the variable cost additives are added to only one component, while no variable costs accrue for all others.}
\label{fig: operational optimization analysis}
\end{figure}

Figure \ref{fig: operational optimization analysis} illustrates the number of hours with \ac{USC} in the system (scatter plots, right ordinate). It further shows \ac{FLH} of the \ac{H2} fuel cell for varying variable costs of the renewable generators or \ac{H2} storage components (lines, left ordinate). 

Adding variable cost to any class of storage components or all renewables can successfully remove \ac{USC} beyond a certain threshold that depends on the solver accuracy. The observed occurrence of \ac{USC} spatially averaged overall modelled nodes with variable cost below $10^{-3}$ is roughly 5200, regardless the level of solver accuracy. Note that for the determination of the \ac{USC} occurrences, any simultaneous charging and discharging below the energy value of 1 MWh was not counted as \ac{USC} to ensure that only significant \ac{USC} energy volumes are considered. Otherwise, \ac{USC} would occur in almost every time step of every scenario with marginal energy volumes, which may be caused by the solver tolerance and the non-uniqueness of the optimal operation of storage assets \cite{Grubel2020OnADMM}. 

More importantly, the \ac{USC} energy volume decreases for increasing cost additives, irrespective of the \ac{USC} occurrence counting method. This decrease is illustrated by the \ac{FLH} curves in Figure \ref{fig: operational optimization analysis}. In general, the \ac{FLH} curves reveal that adding variable costs affects the operation of the storage system. The decline in \ac{FLH} as the variable cost increase is due to two overlapping effects, indicating a trade-off between the removal of \ac{USC} and an undistorted operation of the storage system. For the lower range of the investigated variable cost additive scenarios, the reduction of the \ac{USC} energy volume is the prime driver of the \ac{FLH} decline. In contrast, very high cost additives render the storage system's operation less economically viable, strongly decreasing its optimal use. In the medium-to-high range of the cost additives, both \ac{USC} is prevented and storage operation remains largely constant and undistorted, indicated by the plateau of the \ac{FLH} curves.

In our stylised setting at medium solver accuracy -- which refers to the PyPSA-Eur default values \cite{PyPSA-Eur2021GithubPyPSA-Eur} -- \ac{USC} is fully removed in any considered scenario with an variable cost threshold of at least 10 \euro/MWh or 1 ct/kWh. While for scenarios with a variable cost additive of 1 \euro/MWh \ac{USC} still occurs, the \ac{USC} energy distortions are only marginal with a slight increase of the fuel cell's \ac{FLH} of 21h compared to a cost additive scenario with 10 \euro/MWh (1265h - 1244h). Hence, a variable cost additive of slightly above 1 \euro/MWh (or 0.1 ct/kWh) is likely to prevent \ac{USC} distortions in our case study.

The very cost additive threshold that removes \ac{USC} depends on the level of solver accuracy. In the case of low solver accuracy, the threshold is relatively high at 100 \euro/MWh. However, \ac{FLH} curves hardly stabilise in a plateau, which would indicate that storage operation remains unaffected. This makes it difficult to identify the optimal cost additives that prevent \ac{USC}. Further, such a high variable cost additive level is implausible for \ac{VRE} or storage components. For medium and high solver accuracy, \ac{FLH} curves form a plateau, with the lower end at 10 and 1 \euro/MWh, respectively. Note that we discretely increment the variable cost additives by one order of magnitude. The true underlying thresholds may be in between these increments.



The observed impacts of \ac{USC} on the operation are extreme by design of this study. The operational distortions are amplified through the exclusion of dispatchable renewable and conventional generators, such as biomass, nuclear, or green gas. Including such dispatchable generators may decrease the \ac{USC} energy distortions in many energy models, as this would introduce additional variable costs that reduce the impact of \ac{USC} (see Section \ref{Sec:removing usc}). Nevertheless, this study reveals also that assuming no variable cost for generators or storage technologies, as done in multiple if not most energy system modelling studies \cite{Kittel2021RenewableModeling, Poncelet2016ImpactModels, Welsch2014IncorporatingIreland, deSisternes2016TheSector, Ruhnau2022StorageVariability}, risks unintended operational distortions. If this distortions as well as variable cost additives also impact the investment optimization is investigated in the next section.


\subsection{Effects on investment optimization}

\begin{figure}[p]
\centering
\includegraphics[trim={2.2cm 4.5cm 2cm 0cm},clip,width=0.475\textwidth]{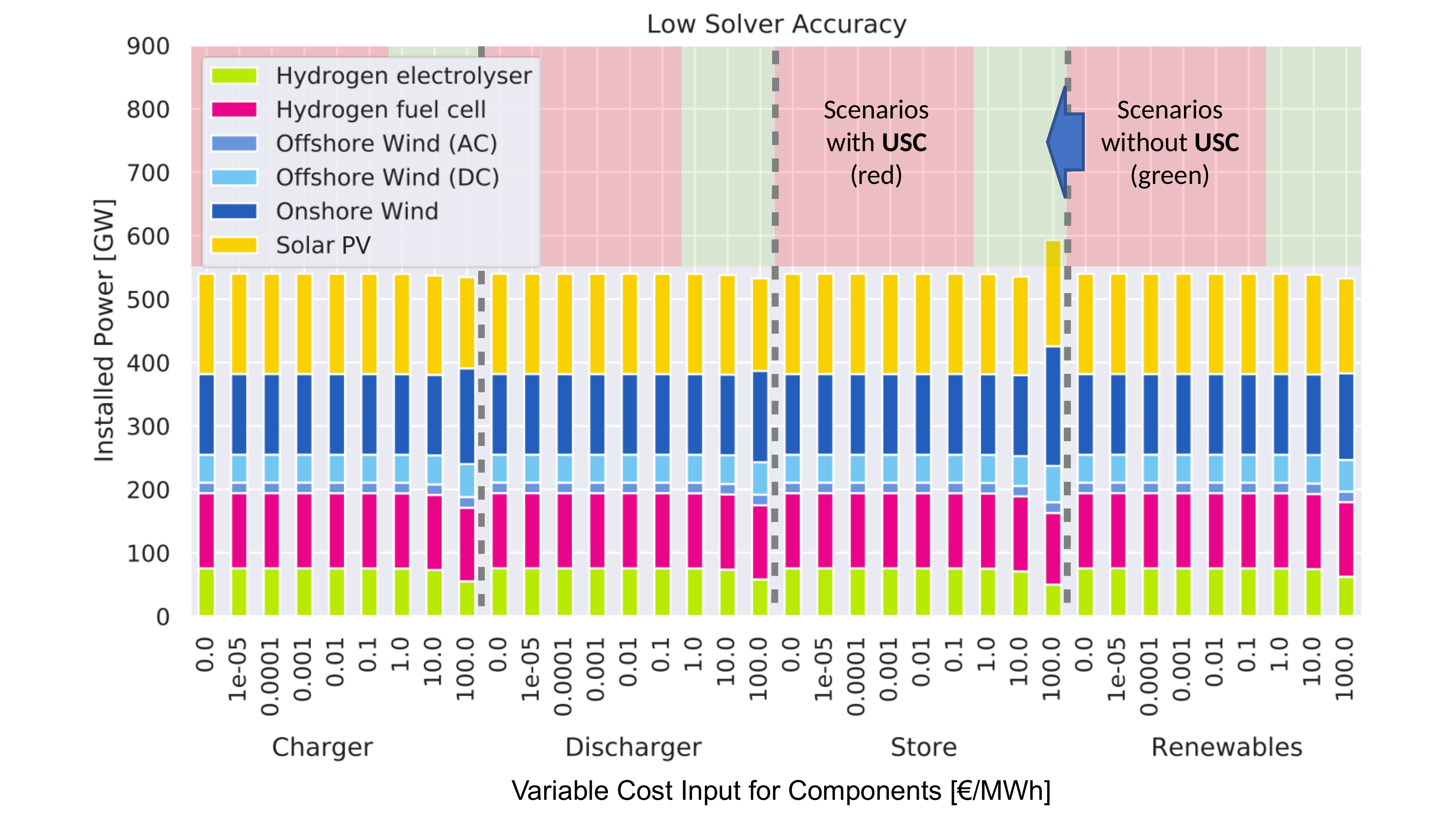}
\includegraphics[trim={2.1cm 6.8cm 3.5cm 0cm},clip,width=0.48\textwidth]{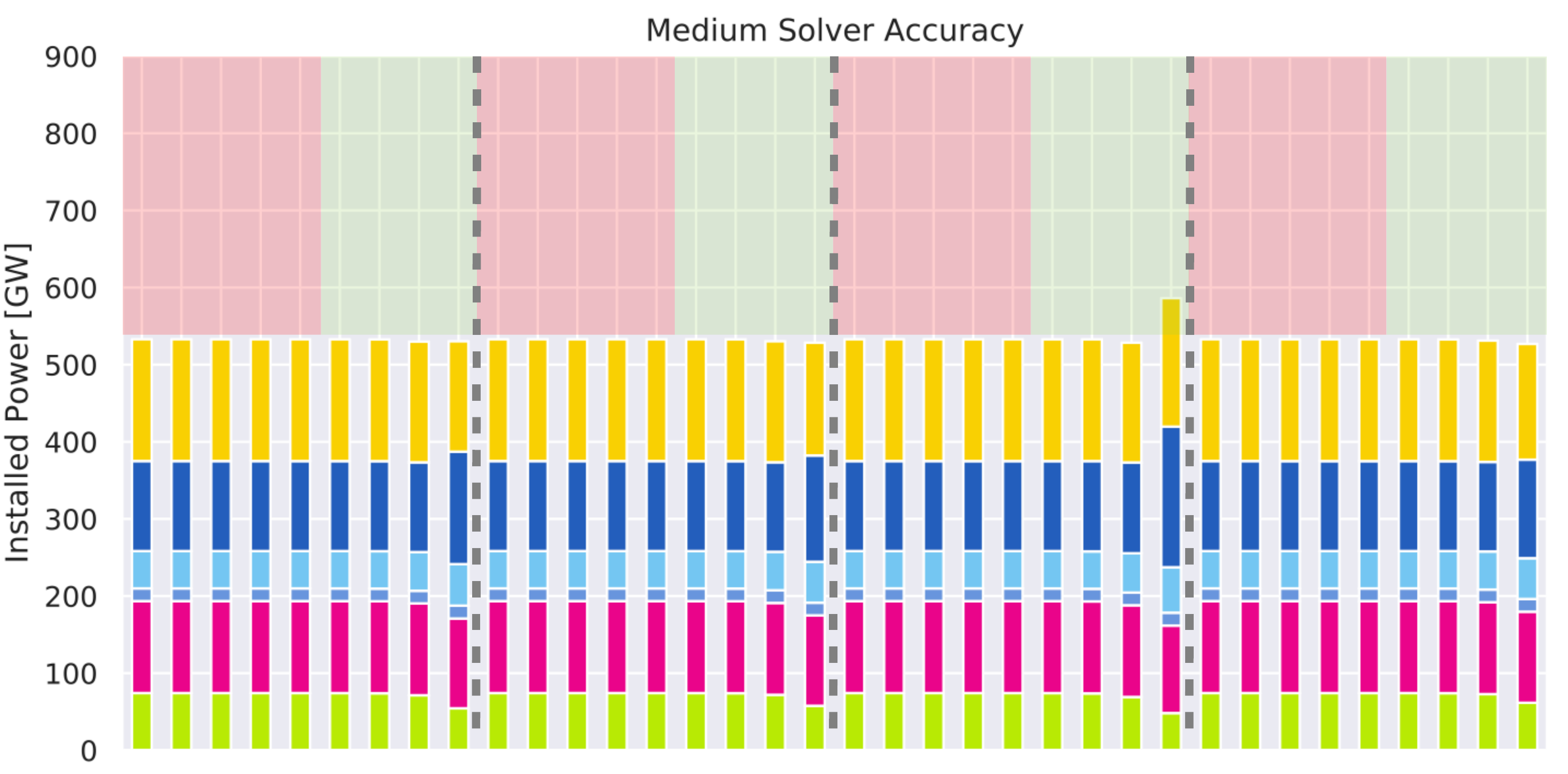}
\includegraphics[trim={2.1cm 0cm 1.8cm 0cm},clip,width=0.48\textwidth]{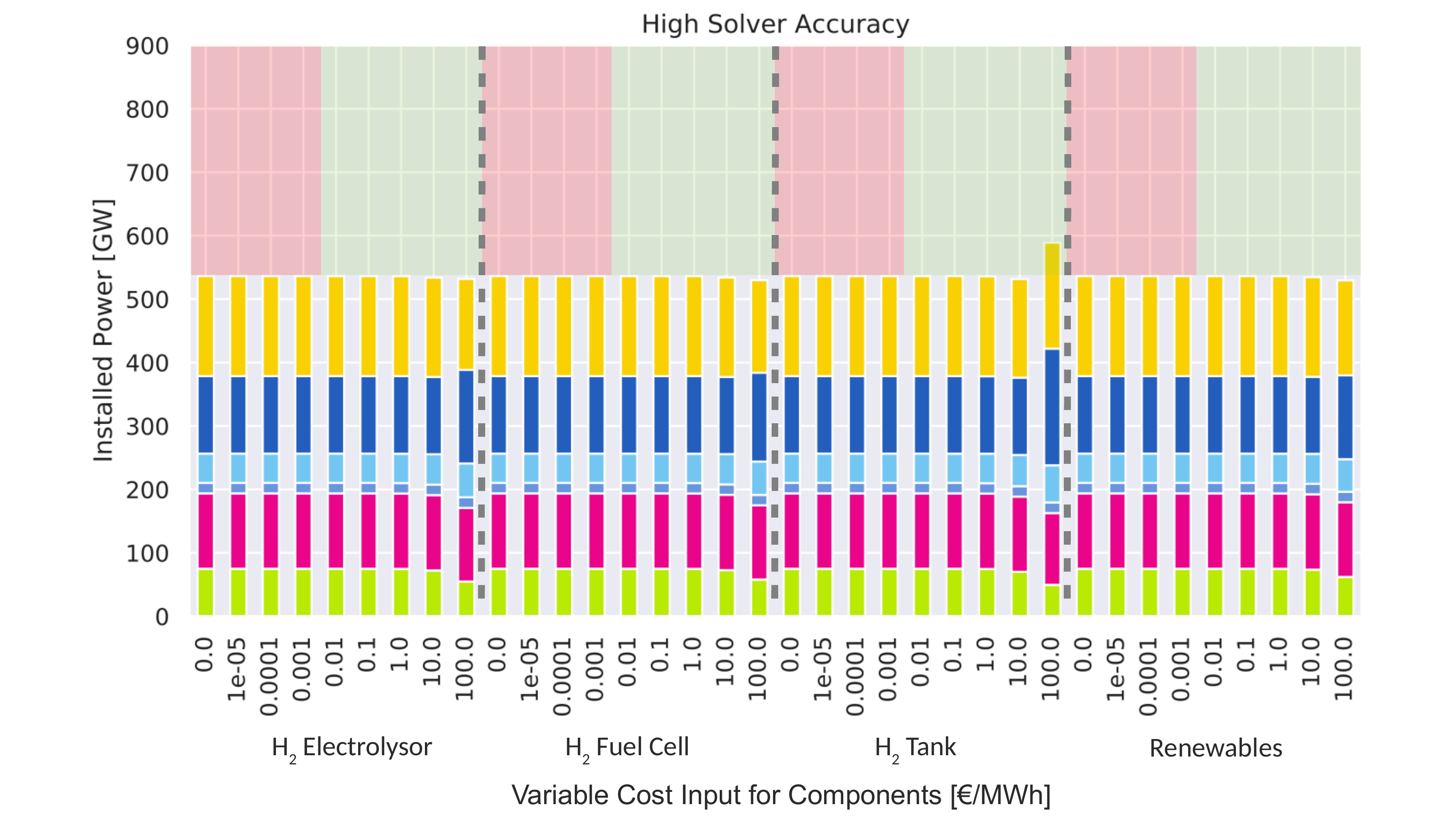}
\caption{Installed capacity of all generation and storage assets for different variable cost additive scenarios. In scenarios in red \ac{USC} arises, while in the the green the effect is prevented. We omit illustrating results from the scenarios using variable costs additives of 1000 \euro/MWh to keep Figure \ref{fig:optimized capacity} and \ref{fig:total system cost} consistent and readable.}
\label{fig:optimized capacity}
\end{figure}

Figure \ref{fig:optimized capacity} illustrates the optimized generation and storage capacity of all modelled scenarios. The optimal installed capacity is not impacted by variable cost additives, irrespective the \ac{USC} occurrence, unless for very high additives of 10 to 100 \euro/MWh. 

In these cases, optimal storage use and capacity decrease and \ac{VRE} use and capacity increase in scenarios that impose additional variable storage costs. This is because storing energy becomes very expensive, rendering more generation capacity optimal to reduce situations that require energy storage operation. 

In scenarios with very high \ac{VRE} variable cost additives, wind power more is used more, while both \ac{PV} and storage are used less. This is due to the more stable generation pattern of wind power compared to solar \ac{PV} which requires more storage to smooth its diurnal generation profile. Additionally, even though variable costs are added only to generator operations, the use of storage becomes more expensive as storage efficiency losses multiplies \ac{VRE} generation cost. For instance, at an electricity price of 100 \euro/MWh and a round-trip efficiency of 25\%, discharging 1 MWh comes at energy procurement costs of 4 MWh $\times$ 100 \euro/MWh (4 MWh must be charged to generate 1 MWh of storage output). This multiplication effect of generation costs in energy storage components would be less of an issue if generation had lower costs. For instance, consider generation costs of 0 \euro/MWh for the same efficiency as before. The effective operation costs of charging and discharging were zero - yet, causing \ac{USC}.

For plausible variable cost additives, \ac{USC} does not impact optimal investment decisions. This is one major difference to \ac{USC} arising in energy models with a renewable energy constraint, where the artifact causes complex distortions of optimal investments \cite{Kittel2021RenewableModeling}.

\subsection{Effects on total system costs}

\begin{figure}[p]
\centering
\includegraphics[trim={4cm 3.8cm 3.6cm 0.3cm},clip,width=0.475\textwidth]{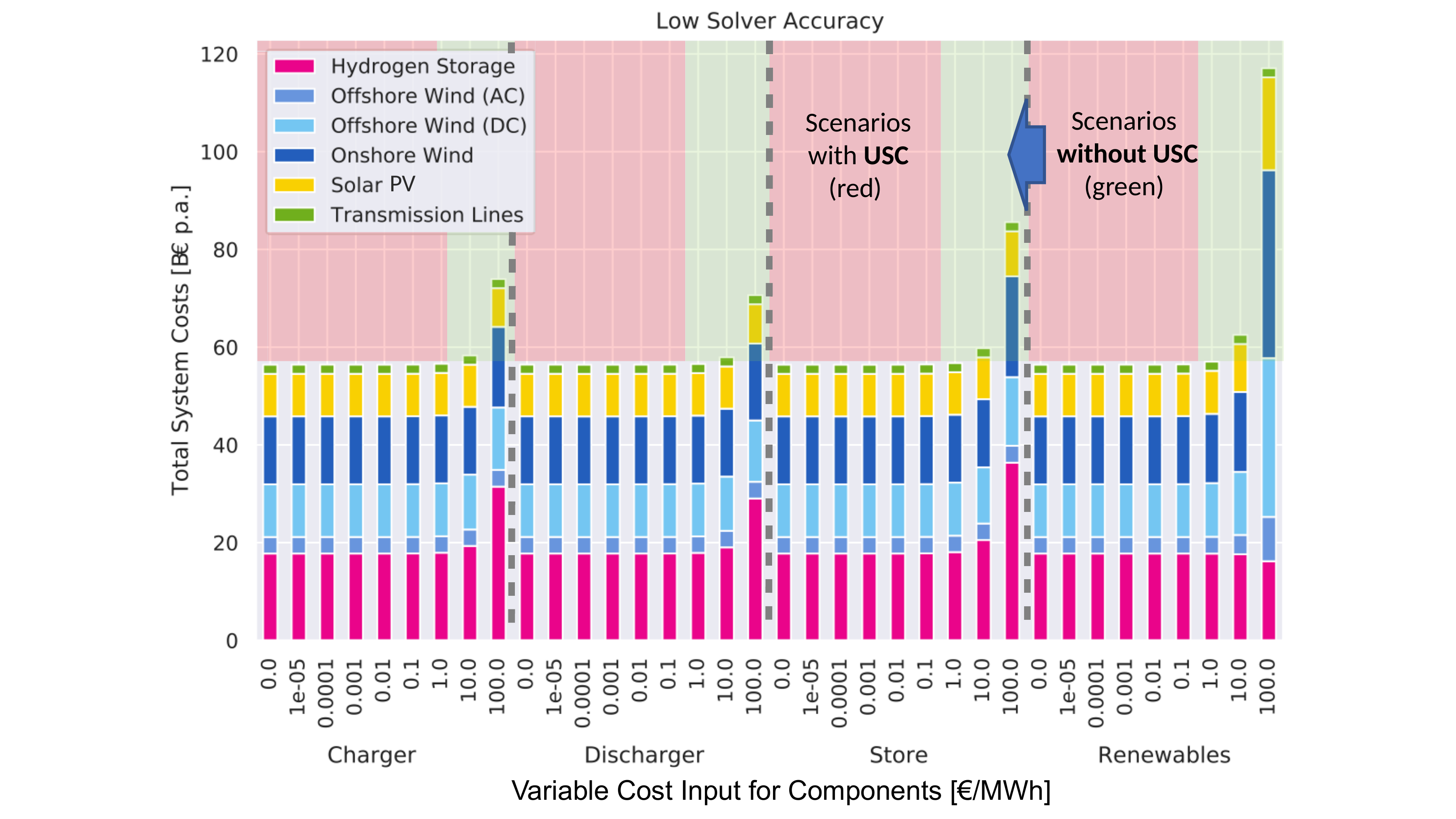}
\includegraphics[trim={3.9cm 6.5cm 5cm 0cm},clip,width=0.48\textwidth]{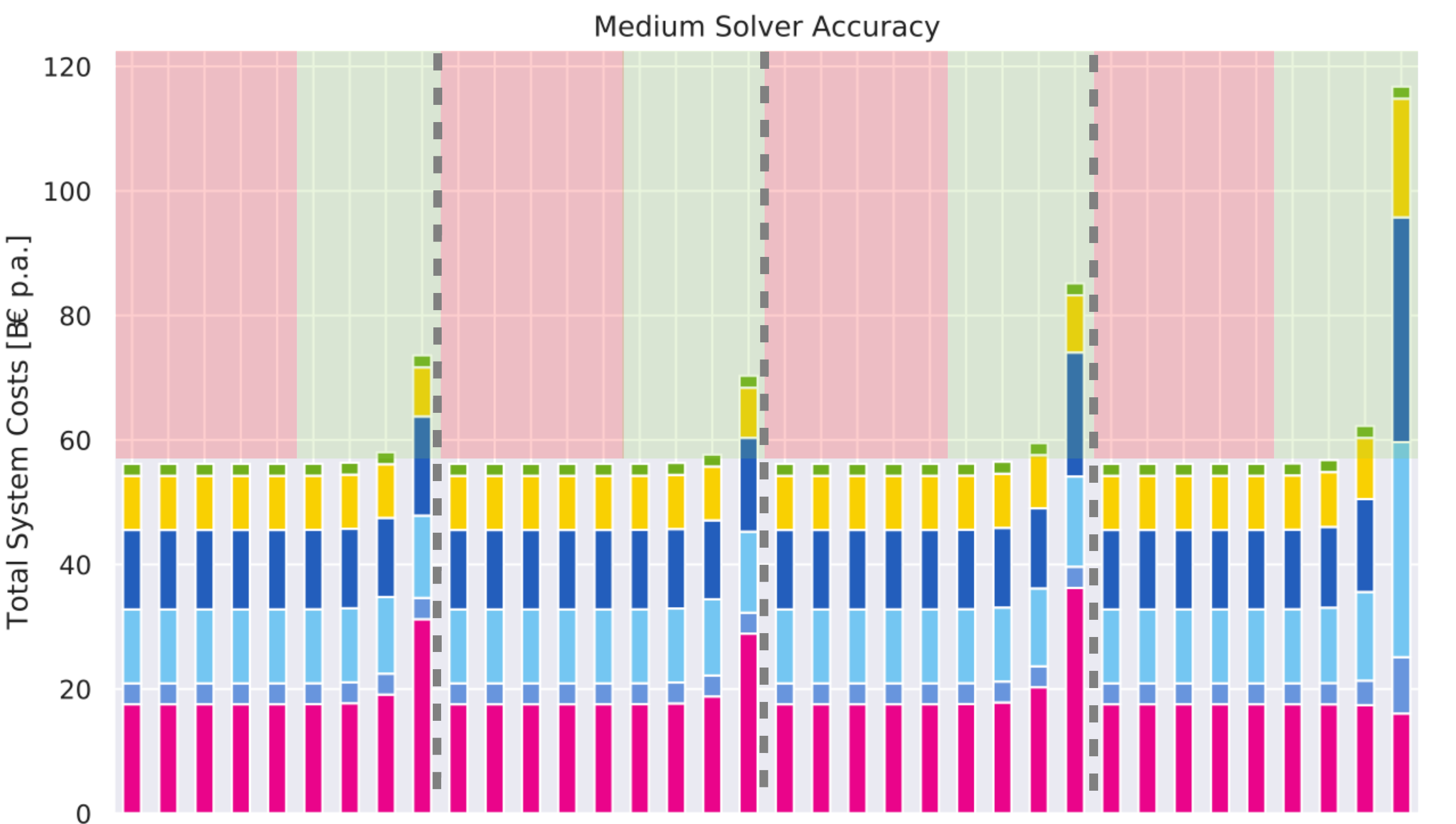}
\includegraphics[trim={3.9cm 0cm 3.3cm 0cm},clip,width=0.48\textwidth]{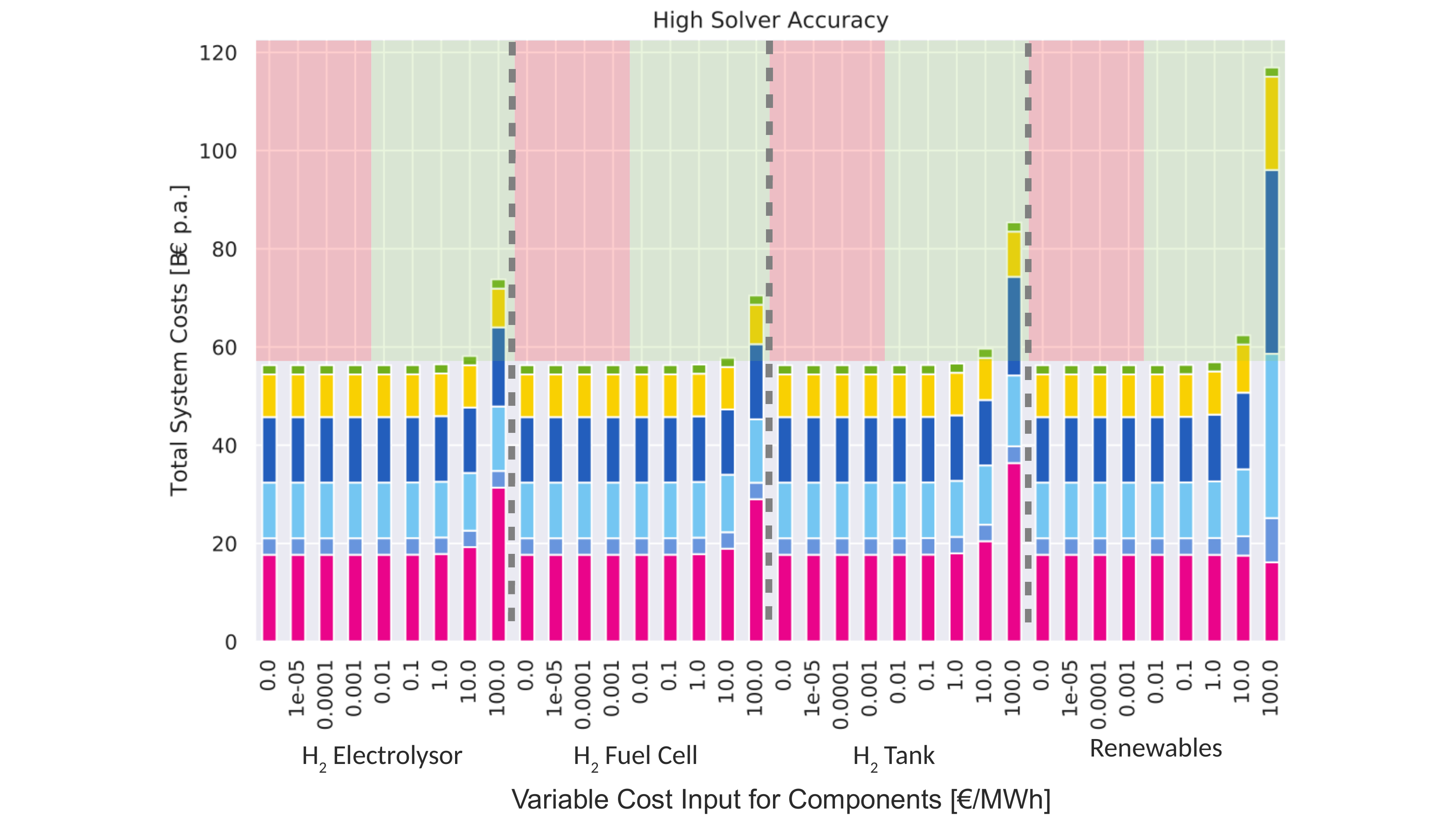}
\caption{Total system costs for different variable cost additive scenarios. In scenarios in red \ac{USC} arises, while in the the green the effect is prevented. Costs of the hydrogen storage consist of electrolyser, fuel cell and H2 storage tank. We omit illustrating total system costs from the scenarios using variable costs additives of 1000 \euro/MWh for readability.}
\label{fig:total system cost}
\end{figure}

The total system costs consist of operational and investment cost and is a key parameter to assess the wider energy system. Figure \ref{fig:total system cost} shows the total system cost results for all optimization runs stacked by system components. It reveals that scenarios with \ac{USC} and applied \ac{USC} removal strategies have only negligible impact on the total system costs unless the variable costs are set too high (above or equal to 10 \euro/MWh). Depending on which technology the variable costs are added, a significant cost increase can be detected due to the extra operational cost that needs to be covered. Again, the assumed values, for instance, of 100 \euro/MWh for the dispatch of all included renewables, might not be realistic but illustrate the impact of mistakenly choosing the wrong values.



\subsection{Variable costs suggestions to alleviate unintended storage cycling}
\label{Section. Setting variable costs right}

Our results suggest that variable costs should be carefully set for all assets to guarantee the removal of \ac{USC} while avoiding any distortion of optimal investment and dispatch decisions. In our stylised setting, the lower end of the plateau of the \ac{FLH} curves in Figure \ref{fig: operational optimization analysis} indicate the optimal threshold for an appropriate cost additive. Values below this threshold are not preventing \ac{USC}, while values above this threshold can lead to investment and operational distortions.

Providing variable costs above a minimum threshold is key to avoid \ac{USC} in models without binding renewable targets, even for technologies with zero or near-zero variable costs. For instance, if a solar \ac{PV} plant is given no or too small variable costs, then solvers cannot recognise them. Consequently, these values should be replaced by the minimum value depending on the solver accuracy and used solver. In our stylised setting of PyPSA-Eur, this minimum value is 1 \euro/MWh at the used default Gurobi solver settings of ($BarConvTol = 1e^{-5}$, $FeasibilityTol = 1e^{-6}$), which remove most \ac{USC} effects. While 1 \euro/MWh or 0.1 ct/kWh may be considered as relatively high, our results in Figures \ref{fig: operational optimization analysis}, \ref{fig:optimized capacity}, and \ref{fig:total system cost} show that these little variable costs neither affect the investment nor the operational costs significantly, making it a threshold candidate to alleviate large \ac{USC} distortions.

There are multiple options for removing \ac{USC}. First, setting minimal variable cost only for one storage component such as charger, store, or discharger for all storage technologies in all regions across the network. Second, imposing minimal variable costs to all \ac{VRE} generation types at all locations (see Equations \ref{usceq} and \ref{nousceq}). Third, imposing cost additives to all storage and generator assets with variable costs below the identified minimum threshold. While out of these options the third affects the variable costs of most technologies, the adjustments are consistent and provide additional redundancy.

The fact that some technologies are assumed with zero costs for energy dispatch could be reconsidered. The concept of \textit{variable cost for lifetime reductions} may justify little cost assumptions for non-zero dispatch generators or storage components. 
Suppose a wind turbine operates 100\% and another one only 50\% of all hours of a year. Further, both turbines are equally maintained by contract-based operation and maintenance service providers (often annualised as fix-costs). However, the twice as much operating turbine is likely to experience on average earlier signs of fatigue in the mechanical structures and power electronics \cite{Ziegler2018LifetimeUK,Wang2013TowardOpportunities}. As a result, the turbine with more operating hours was indeed experiencing costs, namely variable costs for lifetime reductions that are associated with the reduced technical lifetime or extra required operation and maintenance. Such variable cost for lifetime reductions cost may be not trivial and vary across technologies. For instance, in the case of thermal-based processes like steam turbines in concentrated solar power plants, a steady rather than fluctuating operation is preferred. This reduces thermal stresses that otherwise shorten the plant lifetime \cite{Benato2016LTE:Operation}. In summary, even though associating variable cost to the technical lifetime is not trivial, these costs are likely to appear in the energy system, which makes the case for setting variable costs greater than zero.

In Table \ref{tab: variable cost list}, we provide a comprehensive list of variable costs suggestions including a minimum threshold for technologies that are considered near-zero or zero variable cost devices \cite{Blazquez2018TheParadox, Denholm2012DecarbonizingStorage}. In general, these listed values contain large uncertainties since they are not provided in great detail in the literature \cite{Mongird2020Grid2020}. The suggested threshold of 1 \euro/MWh has to be taken with caution as it may not apply to all energy models. The threshold is recommended for a specific modelling tool, namely PyPSA-Eur, while using the Gurobi solver with default accuracy of $BarConvTol = 1e^{-5}$, $FeasibilityTol = 1e^{-6}$. If the model formulation, the solver, or the solver parameters differ, then this suggestion may not be valid anymore. So while the ideal threshold may require to be quantified for each model parameterization separately, the values in Table \ref{tab: variable cost list} could serve as as a default starting point for the identification of an appropriate cost additive.

\begin{table}[h]
\caption{Variable cost suggestions for renewables and a set of storage technologies in energy models based on 2030 data.}
\Bstrut
  \centering
  \resizebox{0.48\textwidth}{!}{
  \begin{tabular}{lll}
    \hline
    Technology \Tstrut\Bstrut & variable cost & Source\\
    & [\EUR{}/MWh] &  \Bstrut \\
    \hline
    onshore wind \Tstrut & $1.4$ & DEA  \cite{DanishEnergyAgencyDEA2019TechnologyData}\\
    offshore wind & $2.7$ & DEA  \cite{DanishEnergyAgencyDEA2019TechnologyData}\\
    PV & $0 \rightarrow 1^{*}$ & Clauser \& Ewert  \cite{Clauser2018TheStudy} \\
    CSP$^{a}$ & $2.9$ & Clauser \& Ewert  \cite{Clauser2018TheStudy}  \\
    CSP + Storage & $4$ & Clauser \& Ewert  \cite{Clauser2018TheStudy} \\
    biomass & $6.7$ & Clauser \& Ewert  \cite{Clauser2018TheStudy}  \\
    tidal & $3.1^{d}$ & [-] \\
    wave & $3.0^{d}$ & [-] \\
    geothermal & $5.6$ & Clauser \& Ewert  \cite{Clauser2018TheStudy} \\
    run of river & $3.6^{a,b}$ & EIA \cite{InternationalLCOEforOceanEnergyTechnologyEIA2021Levelized2021} \\
    hydroelectric dams & $3.6^{a,b}$ & EIA \cite{InternationalLCOEforOceanEnergyTechnologyEIA2021Levelized2021} \\  
    pump-hydro storage & $3.6^{a,b}$ & EIA \cite{InternationalLCOEforOceanEnergyTechnologyEIA2021Levelized2021} \\ 
    battery inverter & $6.8^{a}$ & EIA \cite{InternationalLCOEforOceanEnergyTechnologyEIA2021Levelized2021} \\
    battery storage & $13.5^{a,c}$ & EIA \cite{InternationalLCOEforOceanEnergyTechnologyEIA2021Levelized2021} \\ 
    hydrogen electrolyser & $0 \rightarrow 1^{*}$ & Glenk et al. \cite{Glenk2019EconomicsHydrogen} \\
    hydrogen storage tank & $0 \rightarrow 1^{*}$ & Glenk et al. \cite{Glenk2019EconomicsHydrogen}  \\
    hydrogen fuel cell & $0 \rightarrow 1^{*}$ & Glenk et al. \cite{Glenk2019EconomicsHydrogen}  \Bstrut \\
    \hline


\multicolumn{3}{l}{$^*$ Reported below 1\EUR{}/MWh, but set to 1\EUR{}/MWh to avoid \ac{USC} \Tstrut } \\
\multicolumn{3}{l}{$^a$ Interpolated between 2020 and 2035 } \\
\multicolumn{3}{l}{$^b$ Aggregated as hydroelectric devices by EIA } \\
\multicolumn{3}{l}{$^c$ Assumption of cost split: 2/3 store and 1/3 inverter } \\
\multicolumn{3}{l}{$^d$ Assumed similar to offshore wind. Lack of reliable data \cite{OceanEnergySystemsOES2015InternationalTechnologies}. } \\ 
  \end{tabular}
 } 
  \label{tab: variable cost list}
\end{table}







\section{Conclusion}



Reliable energy model results are essential for planning optimal pathways for the energy transition. However, in energy models without a binding renewable energy target, \ac{USC} may distort the operation of an optimized energy system. For instance, the modelling artifact can significantly increase \ac{FLH} of storage and and renewable assets, in our case study up to 23\%. It can be detected in the case of simultaneous storage charging and discharging of the same storage capacity. Since \ac{USC} is technically infeasible for some storage technologies, e.g., lithium-ion batteries, and can lead to significant operational distortions, it should be removed.

An appropriate level of variable costs can remove \ac{USC}, while keeping the problem formulation linear and convex. However, determining this level is not trivial. The optimization solver may not recognise too low variable costs, which then does not guarantee the removal of \ac{USC}. Hence, we recommend to set a minimum variable costs at a certain threshold that depends on the solver accuracy and tolerance. Very high variable cost, on the other side, may prevent \ac{USC} but can also significantly distort the relative cost ratio of available generation and balancing technologies. As a consequence, optimal investment and dispatch decisions may be flawed. To avoid such model distortions, it is essential to set the variable cost carefully and as accurate as possible.

In our case study, the minimum variable cost threshold of 1 \euro/MWh (or 0.1 ct/kWh) removes all significant \ac{USC} effects at default Gurobi solver accuracy settings without impacting the overall optimization significantly. However, this threshold cannot be generalised to all other energy models. Instead, the efficacy needs to be tested for each model and parameterization. This can be done by checking for simultaneous charging and discharging, while maintaining an appropriate level of solver accuracy. 

We provide a selection of recommended variable costs for a set of storage and renewable generation technologies extracted from recent literature including the identified threshold of 1 \euro/MWh as minimum level wherever necessary. While we did not apply the minimum threshold to all storage and renewable generation technologies at the same time, we could already proof in Figure \ref{fig: operational optimization analysis}, \ref{fig:optimized capacity} and \ref{fig:total system cost} that the threshold sufficiently reduces \ac{USC} distortions without changing the optimization result much when applied for either one of the storage components or all generators. Thus, the values in Table \ref{tab: variable cost list} should be taken with caution. Nonetheless, the variable cost suggestion may serve as a starting point for \ac{USC} tests in other model applications following the philosophy: little variable costs are better than none to lower the risk of significant \ac{USC} distortions. 



Future work, may analyse the removal of other forms of unintended energy losses beyond \ac{USC} with setting appropriate variable costs. For instance, unintended line cycling manifests by simultaneous sending and receiving of electricity through power lines in the distribution and transmission grid \cite{Kittel2021RenewableModeling,Neumann2020ApproximatingProblems}; and sector-coupled cycling, e.g., by electric energy that is converted to heat by boilers and re-electrified at the same time with organic Rankine Cycle plants. These unintended energy losses may originate from the same issue, namely that missing operational costs make a cost-minimising energy model indifferent between possible options to lose unused energy either by cyclic dissipation or \ac{VRE} curtailment. Further, while this study investigates \ac{USC} in energy models triggered by insufficiently specified cost assumptions, \ac{USC} also arises when using additional constraints, e.g., binding renewable energy targets \cite{Kittel2021RenewableModeling}. Constraint-based unintended energy losses are not yet fully explored and merit future research.

\section*{Code and Data availability} 
Code and data to reproduce results and illustrations are available on {\color{blue}\href{https://github.com/pz-max/unintended-storage-cycling}{GitHub}} https://github.com/pz-max/unintended-storage-cycling.

\section*{Credit authorship contribution statement} 
Conceptualisation: M.P., M.K.; 
Methodology: M.P., M.K.; 
Software: M.P., M.K.; 
Validation: M.P., M.K.; 
Formal analysis: M.P., M.K.; 
Investigation: M.P., M.K.; 
Ressources: M.P., M.K., A.K.; 
Data Curation: M.P.; 
Writing - Original Draft: M.P., M.K.; 
Writing - Review \& Editing: M.P., M.K., D.F., A.K.; 
Visualisation: M.P., M.K.; 
Supervision: M.P., D.F., A.K.; 
Project administration: M.P., D.F., A.K.; 
Funding acquisition: M.P., D.F., A.K.; 


\section*{Declaration of Competing Interest}
The authors declare that they have no known competing financial interests or personal relationships that could have appeared to influence the work reported in this paper.

\section*{Acknowledgements} 
This research was supported by UK Engineering and Physical Sciences Research Council (EPSRC) grant EP/P007805/1 for the Centre for Advanced Materials for Renewable Energy Generation (CAMREG) and EPSRC DISPATCH (EP/V042955/1). M.P. would like to thank Fabian Neumann, Harry Van Der Weijde, and Lukas Franken for helpful comments and inspiring discussion.

\mbox{}
\nomenclature[C]{\(c\)}{Specific investment cost (\euro/MW)}
\nomenclature[C]{\(e\)}{Stored energy (MWh)}
\nomenclature[C]{\(\eta\)}{Efficiency}
\nomenclature[C]{\(F\)}{Transmission line capacity (MW)}
\nomenclature[C]{\(g\)}{Generated energy (MWh)}
\nomenclature[C]{\(G\)}{Generator capacity (MW)}
\nomenclature[C]{$h^+$}{Storage charge (MWh)}
\nomenclature[C]{$h^-$}{Storage discharge (MWh)}
\nomenclature[C]{$H^+$}{Storage charge capacity (MW)}
\nomenclature[C]{$H^-$}{Storage discharge capacity (MW)}
\nomenclature[C]{$H_{i,s}^{store}$}{Store capacity (MWh)}
\nomenclature[A]{\(i\)}{Location}
\nomenclature[A]{\(l\)}{Line number}
\nomenclature[C]{\(o\)}{Variable cost (\euro/MWh)}
\nomenclature[A]{\(r\)}{Generator technology}
\nomenclature[A]{\(s\)}{Storage technology}
\nomenclature[A]{\(t\)}{Time step}
\nomenclature[A]{\(T\)}{optimization period}
\nomenclature[C]{\(\overline{T}\)}{Energy to discharging power ratio (MWh/MW)}
\nomenclature[A]{$w_{t}$}{Time weighting}

\nomenclature[C]{\(w\)}{Weighted duration}
\nomenclature[B]{\(FLH\)}{Full load hours (h)}
\nomenclature[B]{\(VRE\)}{Variable renewable energy sources}
\nomenclature[B]{\(USC\)}{Unintended storage cycling}
\nomenclature[B]{\(CSP\)}{Concentrated solar power}
\nomenclature[B]{\(GHG\)}{Greenhouse gas}
\printnomenclature






\bibliographystyle{elsarticle-num-names}
\bibliography{references.bib}







\end{document}